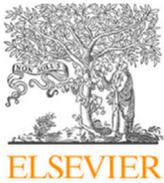
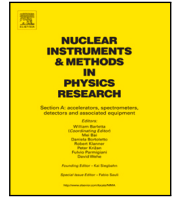

# PICOSEC-Micromegas Detector, an innovative solution for Lepton Time Tagging

A. Kallitsopoulou [a,b],*, R. Aleksan [a,b], Y. Angelis [c,d,1], S. Aune [b], J. Bortfeldt [e], F. Brunbauer [f], M. Brunoldi [g,h], E. Chatzianagnostou [c,d], J. Datta [i], D. Desforge [b], G. Fanourakis [j], D. Fiorina [g,h,2], K.J. Floethner [f,k], M. Gallinaro [l], F. Garcia [m], I. Giomataris [a,b], K. Gnanvo [n], F.J. Iguaz [b,3], D. Janssens [f], M. Kovacic [o], B. Kross [n], P. Legou [b], M. Lisowska [f,a], J. Liu [p], M. Lupberger [k,q], I. Maniatis [c,d,4], J. McKisson [n], Y. Meng [p], H. Muller [f,q], E. Oliveri [f], G. Orlandini [f,r], A. Pandey [n], T. Papaevangelou [a,b], M. Pomorski [s], L. Ropelewski [f], D. Sampsonidis [c,d], L. Scharenberg [f], T. Schneider [f], E. Scorsone [s], L. Sohl [b,5], M. van Stenis [f], Y. Tsipolitis [t], S.E. Tzamarias [c,d], A. Utrobicic [u], I. Vai [g,h], R. Veenhof [f], P. Vitulo [g,h], X. Wang [p], S. White [v], W. Xi [n], Z. Zhang [p], Y. Zhou [p]

[a] *Université Paris-Saclay, F-91191, Gif-sur-Yvette, France*
[b] *Commissariat à l'énergie atomique et aux énergies alternatives (CEA)-IRFU, F-91191, Gif-sur-Yvette, France*
[c] *Department of Physics, Aristotle University of Thessaloniki, University Campus, GR-54124, Thessaloniki, Greece*
[d] *Centre for Interdisciplinary Research and Innovation(CIRI-AUTH), GR-57001, Thessaloniki, Greece*
[e] *Department for Medical Physics, Ludwig Maximilian University of Munich, Am Coulombwall 1, 85748, Garching, Germany*
[f] *European Organization for Nuclear Research(CERN), Geneve 23, 1211, Geveve, Switzerland*
[g] *Dipartimento di Fisica, Università di Pavia, Via Bassi 6, 27100, Pavia, Italy*
[h] *INFN Sezione di Pavia, Via Bassi 6, 27100, Pavia, Italy*
[i] *University of Physics and Astronomy, Stony Brook University, 11794-3800, NY, USA*
[j] *Institute of Nuclear and Particle Physics, NCSR Demokritos, GR-15341, Agia Paraskevi, Attiki, Greece*
[k] *Helmholtz-Institut fur Strahlen-und Kernphysik, University of Bonn, Nißallee 14-16, 53115, Bonne, Germany*
[l] *Laboratório de Instrumentação e Física Experimental de Partículas (LIP), Lisbon, Portugal*
[m] *Helsinki Institute of Physics, University of Helsinki, Fl-00014, Helsinki, Finland*
[n] *Jefferson Lab, 12000 Jefferson Avenue, VA 23606, Newport News, USA*
[o] *Faculty of Electrical Engineering and Computing, University of Zagreb, 10000, Zagreb, Croatia*
[p] *State Key Laboratory of Particle Detection and Electronics, University of Science and Technology of China, 230026, Hefei, China*
[q] *Physikalisches Institut, University of Bonn, Nußallee 12, 53115, Bonn, Germany*
[r] *Friedrich-Alexander-Universität Erlangen-Nurnberg, Schloßplatz 4, 91054, Erlangen, Germany*
[s] *CEA-List, Diamond Sensors Laboratory, CEA-Saclay, F-91191, Gif-sur-Yvette, France*
[t] *National Technical University of Athens, Athens, Greece*
[u] *Ruder Boskovic Institute, Bijenicka cesta 54., 10000, Zagreb, Croatia*
[v] *University of Virginia, VA, USA*

## ARTICLE INFO

## ABSTRACT

The PICOSEC-Micromegas (PICOSEC-MM) detector is a novel gaseous detector designed for precise timing resolution in experimental measurements. It eliminates time jitter from charged particles in ionization gaps by using extreme UV Cherenkov light emitted in a crystal, detected by a Micromegas photodetector with an appropriate photocathode. The first single-channel prototype tested in 150 GeV/c muon beams achieved a timing resolution below 25 ps, a significant improvement compared to standard Micropattern Gaseous Detectors (MPGDs). This work explores the specifications for applying these detectors in monitored neutrino beams for the ENUBET Project. Key aspects include exploring resistive technologies, resilient photocathodes, and scalable electronics. New 7-pad resistive detectors are designed to handle the particle flux. In this paper, two potential scenarios are briefly considered: tagging electromagnetic showers with a timing resolution below 30 ps in an electromagnetic calorimeter as well as individual particles (mainly muons) with about 20 ps respectively.

* Corresponding author at: Commissariat à l'énergie atomique et aux énergies alternatives (CEA)-IRFU, F-91191, Gif-sur-Yvette, France.
  *E-mail address:* alexandra.kallitsopoulou@cea.fr (A. Kallitsopoulou).
[1] Now at European Organization for Nuclear Research (CERN), 1211 Geneve 23, Switzerland.
[2] Now at Gran Sasso Science Institute, Viale F. Crispi, 7 67100 L'Aquila, Italy.
[3] Now at SOLEIL Synchrotron, L'Orme des Merisiers, Departmentale 128, 91190 Saint Aubin, France.
[4] Now at Department of Particle Physics and Astronomy, Weizmann Institute of Science, Hrzl st. 234 Rehovot, 7610001, Israel.
[5] Now at TUV NORD EnSys GmbH & Co. KG.

## 1. Introduction

Managing pile-up density is one of the many challenges facing future physics experiments [1]. Pile-up density, distributed over small space, can significantly complicate event reconstruction. Thus, high-rate experiments demand precise timing to achieve resolution within tens of picoseconds. Integrating timing information, allows for improved triggering and event selection processes. These demanding requirements push existing detector designs to their limits, requiring significant advancements. The objective is to develop detectors with a large area coverage, capabilities of withstanding high radiation fluence and support multi-channel readouts. We propose significant advancements to enhance detector performance. The PICOSEC Micromegas (PICOSEC-MM) drastically improves the timing of standard MicroPattern Gaseous Detectors (MPGDs), by three orders of magnitudes, achieving timing resolution in the tens of picoseconds [2].

To avoid the time jitter produced by charged particles in ionization gaps, the PICOSEC-MM detector configuration exploits extreme UV Cherenkov light emitted in a thin crystal (3 mm) which is then detected by a Micromegas photodetector using CsI photocathode. The detector operates as a double-stage amplification structure, with a gas mixture of $Ne:C_2H_6:CF_4$ (80:10:10). The first single channel prototype achieved remarkable time resolution below 25 ps for Minimum Ionizing Particles (MIPs). The operation principle of such detectors is described in detail in [2,3].

## 2. Forward lepton tagging

### 2.1. ENUBET context

Clean reconstruction of events and strong reduction of the mixing of different events due to pile-up, is also a critical issue for ENUBET. The ENUBET (Enhanced NeUtrino BEams from kaon Tagging) project aims to produce a narrow band beam at the GeV energy scale, having superior control of the neutrino flux, flavor, and energy of the neutrinos produced at the source. The layout of the ENUBET beamline comprises a proton beamline hitting on a target that produces kaons and pions, the transfer line, and the instrumented decay tunnel, large enough to enable the decay of $K^+, \pi^+$. Decay tunnel neutrinos are monitored by the associated lepton in a layered calorimeter. In the direction of the transferline, non-interacting protons are stopped by the proton dump. Finally in the tunnel exit hadron dump stops all particles but neutrinos. A full description of the ENUBET beamline is described in [4].

In this facility there is no one to one correlation between positrons in the beamline and neutrinos in the far detector, so a possible sub-ns sampling given by the PICOSEC-MM detector can give this correlation on an event-by-event basis, to determine the flavor of the neutrino. To monitor the low-energy $\nu_\mu$ component, muon stations are planned to be installed in the hadron dump enabling measurement of the muon range resulting from pion decay [5]. A graphical representation of this scenario is described in Fig. 1. The PICOSEC-MM technology can also be used embedded in the electromagnetic calorimeter (EMC), for precise timing of electron showers, and/or as a T0 timing layer at the ENUBET Tagger for single MIP detection.

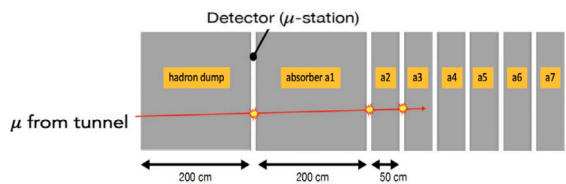

**Fig. 1.** Schematic of the muon stations and absorbers configuration to be installed as the hadron dump. The gray slabs represent the absorbers (made out of iron or rock) while the white slices, 8 in total, are the muon detector planes, where the PICOSEC-MM detector can be placed [5].

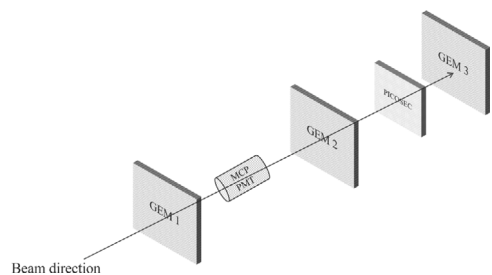

**Fig. 2.** Schematic of the experimental setup for time response test of the multipad detector to 150 GeV/c muons.

#### 2.1.1. Detector description and testing

Given the high rate of events expected in ENUBET, integrating resistive technology into the anode structure of the PICOSEC-MM prototypes will offer distinct advantages in its stable operation. One of the key benefits is the elimination of the destructive effects caused by discharges, as well as the detector stability in harmful environments caused by high-energy particles, such as pions, [6]. The goal of those prototypes will be to exploit these advantages while maintaining precise timing resolution.

For this specific proof of concept, new 10 M Ohm/□ ($cm^2$) resistive detectors were designed. The MM board was fabricated as $4 \times 0.035$ mm thick FR4 outer layers, while the readout structure was arranged in 7 hexagon pads of 1 cm diameter. The signal from each pad was routed to the SMA connectors on the top of the readout board.

The time response of the 7-pad detector to 150 GeV muons was measured at the CERN SPS H4 secondary beamline. The beam size is 8 cm diameter, corresponding to approximately $10^5$ muons/spill, and a measured rate of few kHz/$cm^2$. The experimental setup (as seen in Fig. 2) included a triple GEM detector telescope to provide particle hit information and MCP-PMTs for timing reference or triggering. In the readout chain configuration, an induced signal from the detector is amplified using custom-made 10-channel, 38 dB, 650 MHz amplifier cards with built-in discharge protection, inspired by the RF amplifier design of [7]. The final digitization and sampling of 10 GS/s is done by LeCroy WR8104 oscilloscope. The same oscilloscope is used for the acquisition of the reference time detector signal and serial bit-stream with event ID obtained from the tracker.

The first timing measurements were made with a prototype with drift gap thickness of 150 μm and a CsI (18 nm) photocathode on a thin Cr layer (3.3 nm). The arrival time of PICOSEC-MM signals is determined by analyzing the fully digitized electron peak waveform (E-peak), i.e. by fitting a logistic function to the leading edge of the E-peak. The Signal Arrival Time (SAT) was then defined as the 20% of the

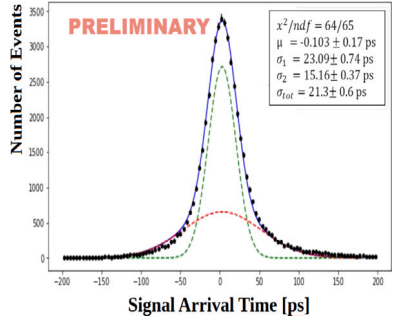

**Fig. 3.** Signal Arrival Time distribution of the central Pad-0 of the multipad detector, operating in the highest gain of 550 V on the cathode and 275 V on the anode after time walk corrections, resulting in a timing resolution of 21.3 ± 0.6 ps. This number refers to approximately 11 photoelectrons per muon track.

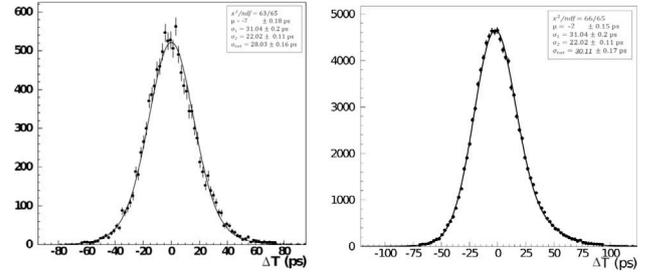

**Fig. 4.** Signal Arrival Time distributions of the central Pad-0 of the multipad detector, after time walk corrections. (a) The timing response on electrons considered as Minimum Ionizing Particles (MIPs) is 28.03 ± 0.16 ps. (b) The timing response on muons as MIPs, in the same gain configuration, is 30.11 ± 0.17 ps. Those numbers refer to approximately 4 photoelectrons per muon track.

E-peak maximum relative to the time-reference device signal. Finally, the detector's timing resolution was defined as the RMS of the SAT distribution. The best response of the central pad of the 7pad detector was 21.3 ± 0.6 ps, Fig. 3.

### 2.2. Detector response on ENUBET's requirements

#### 2.2.1. Tagging individual leptons

Following the successful proof of concept for the newly designed prototype, which demonstrated a timing resolution below 25 ps for single MIPs, the focus has now shifted to timing electromagnetic showers. The measurements were performed again on the CERN SPS H4 secondary beamline, using 30 GeV electrons. The beam size is 2.5 cm in diameter, given a measured rate of the order of a few MHz/cm$^2$.

During previous measurements at CERN SPS, the detector was subjected to a high pion flux. It was observed aging of the CsI photocathode caused by significant ion backflow within the detector. This indicated that CsI is unsuitable for environments with high particle flux [8]. More recent studies in particle beams have shown that photocathodes based in boron carbide material deposition (B4C) can be a robust alternative [9].

The timing measurements for PICOSEC-MM response in electron beam, were made with the same detector prototype with drift gap thickness of 150 μm and a B4C (12 min deposition time) photocathode on a thin Cr layer. The purpose of those studies were to prove the stable performance in different leptonic beams, were particles are considered as MIPS. The detector gain had to be readjusted, reducing the pre-amplification voltage to 500 V, while keeping the amplification voltage to the same value of 275 V. This modification was necessary to withstand the higher particle rate, ensuring stable operation of the detector. As can be seen in Fig. 4, the detector response in electrons as MIPs is 28.03 ± 0.16 ps, while for muons as MIPs is 30.11 ± 0.17 ps.

#### 2.2.2. Tagging electromagnetic showers

To test the detector's response to showers we had to apply some modifications to the telescope. For this measurement there is no tracking information from the triple GEM telescope. The setup comprises of the beam first passing through the MCP-PMT reference device, that still remain our trigger system, followed by two small scintillators (5 × 5 mm) to monitor that is a single particle beam. Next in the line there is 5 cm iron absorber, the size of which was chosen in order the maximum of the shower to be inside of the material, accompanied by a large (10 × 10 cm) scintillator together with a 5 mm hole veto one, to monitor that the shower has initiated inside the absorber. A schematic and a realistic representation of such setup can be found in Fig. 5.

The electron beam was adjusted to hit perpendicular the center of the detector. The response of the 7pad gain individually on the electromagnetic showers can be seen in Fig. 6, achieving a maximum value of 175 mV.

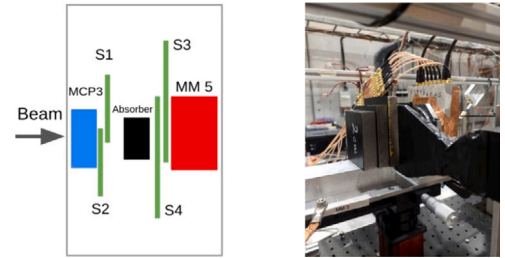

**Fig. 5.** (a) Schematic drawing of electron beam setup. From left with blue the MCP-PMT reference device, with green the two small scintillators (5 × 5 mm), with black the 5 cm iron absorber, with green the large (10 × 10 cm) scintillator together with a 5 mm hole veto one and finally with red the PICOSEC-MM detector. (b) Real representation of the electron beam setup.

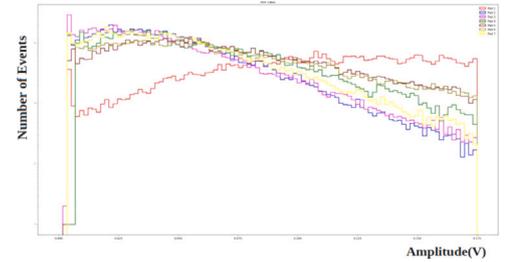

**Fig. 6.** Amplitude distribution of the 7pad individually to the electromagnetic showers. With red is the central Pad-0. All pads were active due to showers with a maximum amplitude value 175 mV.

The detector's response to electromagnetic showers for the central Pad-0 was 20.06 ± 0.1 ps. In Fig. 7 an individual pad response can be seen. Overall, the detector's response to electromagnetic showers was proven to be around 30 ps on average.

### 3. Conclusions

This study successfully demonstrates that the PICOSEC-MM detection concept overcomes the timing limitations commonly associated with gaseous detectors, as proven by our measurements in low-rate environments. We have achieved significant progress in scaling the prototypes while maintaining excellent timing resolution, although further optimization and testing of new methodologies remain necessary. We have identified gain settings as a critical factor affecting both the robustness and resolution of the detector, necessitating careful fine-tuning. Notably, we successfully tested the PICOSEC-MM detector in particle showers for the first time, achieving a timing resolution of 20.06 ± 0.1 ps, result that meets the criteria for the ENUBET application. Additionally, tests to evaluate the radiation hardness and aging of the photocathodes are essential, particularly for this potential application.

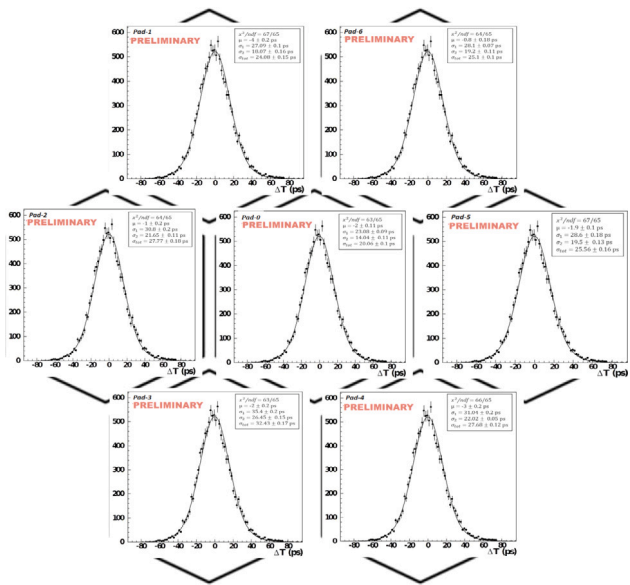

**Fig. 7.** Individual pad, of the 7pad PICOSEC-MM detector, response to electromagnetic showers, created by 5 cm iron absorber.

**Declaration of competing interest**

The authors declare the following financial interests/personal relationships which may be considered as potential competing interests: Alexandra Kallitsopoulou reports financial support was provided by French National Research Agency. If there are other authors, they declare that they have no known competing financial interests or personal relationships that could have appeared to influence the work reported in this paper.

**Acknowledgments**

We acknowledge the financial support of French National Research Agency (ANR), France, project ID ANR-21-CE31-0027; the Cross-Disciplinary Program on Instrumentation and Detection of CEA, the French Alternative Energies and Atomic Energy Commission, France; the PHENIICS Doctoral School Program of Paris-Saclay University, France; the EP R&D, CERN Strategic Program, China on Technologies for Future Experiments; the RD51 collaboration, in the framework of RD51 common project; the Fundamental Research Funds for the Central Universities of China; the Program of National Science Foundation of China (grand number 11935014); the COFUND-FP-CERN-2014 program (grand number 665779); the Fundação para a Ciência e a Tecnologia (FCT), Portugal (CERN/FIS-PAR/0005/2021); the Enhanced Eurotalents program(PCFundação para a Ciência e a TecnologiaFUND, Portugal-GA-2013-600382) and the US CMS program under DOE, USA contact No. DE-AC02-07CH11359.